\documentclass[12pt]{article}
\pdfoutput=1

\usepackage{slashed}

\usepackage{hyperref}
\hypersetup{
  colorlinks = true,
  urlcolor = blue,
  linkcolor= red,
  citecolor=green,
  filecolor=magenta,
}

\usepackage{graphicx}

\usepackage{epstopdf}

\usepackage{pdfpages}

\usepackage{amsfonts}

\usepackage{amssymb}

\usepackage{booktabs}

\newcommand{\ud}{\mathrm{d}}

\title{Bounds on large extra dimensions from the simulation of black hole events at the LHC}

\author{Shaoqi Hou\thanks{Email: shou$@$crimson.ua.edu} , Benjamin Harms\thanks{Email: bharms$@$bama.ua.edu}
\\ \small{Department of Physics \& Astronomy, The University of Alabama}
\and
Marco Cavagli\`a\thanks{Email: cavaglia$@$phy.olemiss.edu}
 \\ \small{Department of Physics \& Astronomy, The University of Mississippi}
}

\date{}

\begin{document}

\maketitle

\abstract{If large extra dimensions exist, the Planck scale may be as low as a TeV and microscopic black holes may be produced in high-energy particle collisions at this energy scale. We simulate microscopic black hole formation at the Large Hadron Collider and compare the simulation results with recent experimental data by the Compact Muon Solenoid collaboration. The absence of observed black hole events in the experimental data allows us to set lower bounds on the Planck scale and various parameters related to microscopic black hole formation for a number ($3-6$) of extra dimensions. Our analysis sets lower bounds on the fundamental Planck scale ranging from 0.6 TeV to 4.8 TeV for black holes fully decaying into Standard Model particles and 0.3 TeV to 2.8 TeV for black holes settling down to a remnant, depending on the minimum allowed black hole mass at formation. Formation of black holes with mass less than 5.2 TeV to 6.5 TeV (SM decay) and 2.2 TeV to 3.4 TeV (remnant) is excluded at 95\% C.L.
 Our analysis shows consistency with and difference from the CMS results.}

\section{Introduction}

The Standard Model of Elementary Particles (SM) is one of the most successful hypotheses in physics \cite{mpp}. However, the SM fails to explain the \emph{hierarchy problem}, i.e., the huge gap between the \emph{electroweak scale} ($M_\mathrm{EW} \sim$1 TeV) and the \emph{observed Planck scale} ($M_\mathrm{Pl}\sim10^{16}$ TeV). The ADD model \cite{add} provides a way to solve the hierarchy problem by introducing a number $n$ of large, compactified spatial dimensions (LEDs). Gravitons can propagate in the $D(=n+4)$-dimensional space-time \emph{bulk}. SM particles are confined to the 4-dimensional \emph{brane}. Assuming compactification on a torus with equal radii $R$, the observed Planck mass $M_\mathrm{Pl}$ is related to the 4-dimensional \emph{fundamental} Planck mass $M_*$ by $M_\mathrm{Pl}^2=(2\pi R)^n M_*^{D-2}$. If $R$ is sufficiently large, the fundamental Planck mass $M_*$ may be as low as a few TeVs. 

If the ADD model is realized in nature, strong gravitational effects should manifest themselves in physical processes at the TeV scale. Gravitational phenomena at the TeV scale could include, for example, graviton and Kaluza-Klein mode production in particle scattering\cite{TLZH} and even microscopic black hole formation \cite{adm}. The Large Hadron Collider (LHC), operating at a center of mass energy of several TeVs, can be used to probe the appearance of these new physical processes and shed light on the existence of large extra dimensions \cite{mpp,grw,giddings,dl,kcheung}.

To date, experimental results have not confirmed the existence of large extra dimensions \cite{exo-12-009, conf-2013-051, proc-2012-262,proc-2013-094, proc-2013-073,proc-2013-069, exo-12-060, exo-12-010}. These null results set upper bounds on the size of the large extra dimensions, or equivalently, lower bounds on the the fundamental Planck scale. Models with one or two large extra dimensions have been ruled out \cite{pdg14,catfish}, such as the Fermi Large Area Telescope \cite{fermi-lat}. Constraints on space-times with three large extra dimensions from astrophysical and cosmological experiments are generally very stringent, although they typically suffer from large systematic errors. The observation of Supernova SN1987A sets a lower limit on $M_D$ of 2.4 TeV for $n=3$ \cite{hanhart}, where the reduced Planck mass $M_D$ is related to $M_*$ by \cite{grw},
\begin{equation}\label{md2mx}
M_D=\Big[\frac{(2\pi)^n}{8\pi}\Big]^{\frac{1}{n+2}}M_*. 
\end{equation}
Neutron star-derived limits constrain $M_D$ to be larger than 76 TeV for $n=3$ \cite{hannestad}. Collider experiments provide less stringent, albeit more accurate limits on $M_D$ for $n\ge3$ from non-observation of perturbative processes. These limits (in units of TeV) are shown in Table \ref{table1}.

\begin{table}[h]

\center

\begin{tabular}{l*{11}{c}}


& \multicolumn{11}{c}{Collider experiments} \\

\cmidrule(lr){2-12}

$n$ & \cite{lep} & \cite{cdf} & \cite{d0-1} & \cite{d0-2}  & \cite{exo-11-059} & \cite{exo-11-058} & \cite{conf-2012-147} & \cite{d87} & \cite{exo-12-048} & \cite{exo-12-031,exo-12-027}  & \cite{cms-exo-12-048} \\

\midrule

3 & 1.20 & 1.15 & 0.86 & 0.80 & 3.21 & 1.25 & 3.16 & 0.98 & 4.29 & 1.20 & 4.77 \\

4 & 0.94 & 1.04 & 0.84 & 0.73 & 2.80 & 1.26 & 2.84 & 0.96 & 3.71 & 1.17 & 3.97 \\

5 & 0.77 & 0.98 & 0.82 & 0.66 & 2.55 & 1.26 & 2.65 & 0.92 & 3.31 & 1.12 & 3.73 \\

6 & 0.66 & 0.94 & 0.80 & 0.65 & 2.36 & 1.29 & 2.58 & 0.88 & 3.12 & 1.07 & 3.53 \\

\bottomrule

\end{tabular}

\caption{95\% C.L.\ lower limits on $M_D$ (TeV) from collider experiments.}\label{table1}

\end{table}

Lower bounds on the Planck scale can also be derived by non-observation of production and decay of TeV BHs in collider experiments and cosmic ray observations \cite{nw,fs,kn,ag,yuehara,krt,amfhhh,afgs}. The  Compact Muon Solenoid (CMS) and A Toroidal LHC ApparatuS (ATLAS) collaborations have conducted searches for BH signatures at the LHC \cite{exo-12-009,atlasjhep}, setting limits on the production cross section and the minimal BH mass $M_\mathrm{min}$, i.e., the minimum mass at which a BH can form. Depending on model assumptions, the CMS (ATLAS) collaboration excludes $M_\mathrm{min}$ below 4.3 to 6.2 TeV (4.8 to 6.2 TeV) at 95\% C.L.  

The purpose of this paper is to revisit and extend the above results from CMS. We simulate production and decay of microscopic BHs at the LHC with the Monte Carlo generator CATFISH (v2.10) \cite{catfish}. We then derive experimental bounds on the fundamental Planck mass and the number of extra dimensions by comparing simulation results with model-independent experimental limits on BH production from the CMS Collaboration \cite{exo-12-009}. The absence of observed black hole events in the CMS experimental data allows us to set bounds on various physical parameters of the ADD model and constrain the minimum mass of TeV-scale black holes that may form in hadronic scattering processes. 

\section{Black Hole Formation in Particle Collisions}

According to the \emph{Hoop Conjecture} \cite{kt}, a BH forms when a mass $M$ is confined into a region of typical size equal to the Schwarzschild radius for that mass, $R_\mathrm{S}(M)$. Therefore, if two particles collide with center of mass energy $\sqrt{s}$ and impact parameter smaller than $R_\mathrm{S}(\sqrt{s})$, a BH may form. If $R_\mathrm{S}\ll R$, as expected in the ADD scenario, the newly formed BH lives in a $D$-dimensional space-time with negligible curvature at the BH scale. In this case, the Schwarzschild radius of the BH can be expressed as \cite{mp,mc,pk},
\begin{equation}\label{rs}\displaystyle
R_\mathrm{S}=\frac{1}{\sqrt{\pi}M_*}\left[\frac{8\Gamma\left(\frac{n+3}{2}\right)}{n+2}\right]^{\frac{1}{n+1}}\left(\frac{M}{M_*}\right)^{\frac{1}{n+1}}\,,
\end{equation}
where $M=(1-y)\sqrt{s}$ and $y$ is the fraction of energy which escapes into the bulk as gravitons, depending on the impact parameter. The Hoop Conjecture implies a BH production cross section $\sigma(s,n,y)=\pi F R^2_\mathrm{S}$, where the form factor $F\le 1$ is related to $y$ and accounts for the energy of the colliding particles that is not trapped in the event horizon, the so called ``graviton energy loss at formation.'' Since BH production in hadron colliders occurs at the parton level, the total cross section for a hadronic collision is obtained by integrating on the Parton Distribution Functions (PDFs) of the hadrons \cite{mc}
\begin{equation}\label{sigmatt}
\sigma_{pp\rightarrow \mathrm{BH}}(s,n,y)=\sum_{ij}\int_0^12z\ud z\int_{x_m}^1\ud x\int_x^1\frac{\ud x\rq{}}{x\rq{}}f_i(x\rq{},Q)f_j(x/x\rq{},Q)\sigma(\sqrt{xs},n,y)\,,
\end{equation}
where $f_i(x,Q)$ are the PDFs with four-momentum transfer squared $Q$ and $z$ is the impact parameter normalized to its maximum value. The cutoff at small $x$ is $x_m={M_\mathrm{min}}^2/\{s[(1-y(z)]^2\}$, where $M_\mathrm{min}$ is the minimum-allowed mass of the BH. The total cross section in absence of graviton energy loss at formation is recovered by setting $F=1$ (Black Disk (BD) cross section) \cite{catfish}. 

If the initial BH mass is much larger than the Planck mass, a semiclassical treatment suggests that the newly-formed BH decays through four, possibly distinct stages: {\it balding}, {\it spin-down}, {\it thermal evaporation} and {\it quantum decay} \cite{gm}. During the balding stage, the BH radiates multipole momenta and quantum numbers \cite{giddings, gl}, eventually settling down to a $D$-dimensional Kerr geometry. Angular momentum is radiated during the spin-down stage \cite{gl}. The Schwarzschild BH then decays into elementary particles through the Hawking mechanism (thermal evaporation stage). Most of the energy of the BH is radiated in this stage, with SM particles dominating the decay products. When the mass of the evaporating BH approaches the Planck scale, $Q_\mathrm{min}\sim M_*$, the BH enters the quantum phase, where the decay ceases to be semiclassical and becomes dominated by quantum gravitational effects. 

\section{BH Event Simulations}

Several Monte Carlo generators for BH production at colliders have been developed over the years: TRUENOIR \cite{truenoir}, CHARYBDIS2 \cite{ch,ch2}, BlackMax \cite{bm,bmm} (used by CMS), QBH \cite{qbh} and an unnamed generator by Tanaka \emph{et al.} \cite{tanaka}. Our analysis is based on CATFISH (\emph{Collider grAviTational FIeld Simulator for black Holes}) \cite{catfish}. CATFISH is a Fortran 77 Monte Carlo generator designed specifically for simulating BH events at CERN's LHC. It incorporates three models for BH formation and cross section: BD, Yoshino-Nambu (YN) Trapped Surface (TS) \cite{yn}, and Yoshino-Rychkov (YR) improved TS model \cite{yr}. The lack of a quantum theory of gravity requires a phenomenological treatment of the final stage. CATFISH offers the choice of simulating the quantum phase by either non-thermally decaying the BH into a number $n_p$ of hard quanta, each with energy $Q_\mathrm{min}/n_p$, or forming a BH remnant. CATFISH also incorporates several other physical effects, such as inelasticity, exact field emissivities and corrections to semiclassical BH evaporation. The generator interfaces to the PYTHIA Monte Carlo fragmentation code \cite{py} using the Les Houches interface \cite{lh}. In our analysis, we run CATFISH (v2.10) with the CTEQ6PDF PDF set and PYTHIA (v6.425) Tune Z1.  

The simulation of a BH event in CATFISH follows these steps. First, CATFISH computes the total and differential cross sections for the BH formation. The initial BH mass is sampled from the differential cross section. The BH is then decayed through the Hawking mechanism until the BH mass reaches the quantum limit, where a final non-thermal hard event is generated or a BH remnant is created. The unstable quanta emitted by the BH are instantaneously hadronized or decayed by PYTHIA, which also simulates initial- and final-state radiation particles. To determine the physics of BH formation and decay, CATFISH uses several external parameters and switches:

\begin{itemize}
\item \emph{ADD parameters}
\begin{enumerate}
\item Fundamental Planck mass: \verb+MSTAR+=$M_*$.
\item Number of extra dimensions: \verb+NEXTRADIM+$=n$, \verb+NEXTRADIM+ = 3, 4, 5, 6.
\end{enumerate}
\item \emph{BH formation parameters}
\begin{enumerate}

\item Graviton energy loss at formation: \verb+GRAVITONLOSS+ = 0 (BD model), 1 (YR or YN TS models, see below). 

\item Gravitational loss model: \verb+GRAVITONMODEL+ = 0 (YN TS model \cite{yn}), 1 (YR improved TS model \cite{yr}).

\item Minimum initial BH mass in Planck units: \verb+XMIN+$=M_\mathrm{min}/M_*$, \verb+XMIN+$\ge$ 1.

\end{enumerate}

\item \emph{BH evaporation parameters}

\begin{enumerate}

\item Minimum quantum BH mass in Planck units: \verb+QMIN+$=Q_\mathrm{min}/M_*$, \verb+QMIN+ $\le$ \verb+XMIN+. 

\item Number of quanta emitted in the Planck phase: \verb+NP+ = $n_p$. When \verb+NP+=0, the BH forms a stable remnant with mass $Q_\mathrm{min}$. 

\end{enumerate}

\end{itemize}

The primary goal of this investigation is to determine lower bounds on the $D$-dimensional fundamental Planck scale for different values of \verb+NEXTRADIM+, \verb+XMIN+ and \verb+NP+ using the model-independent 95\% C.L.\  upper limits on the BH cross section from the CMS search \cite{exo-12-009}. We also derive lower limits on $M_\mathrm{min}$ and \verb+XMIN+ for fixed \verb+MSTAR+, \verb+NEXTRDIM+ and \verb+NP+.  For simplicity, we consider only BH formation with \verb+GRAVITONLOSS+=0 (BD cross section), \verb+XMIN+=\verb+QMIN+ and final decay in 2, 4, or 6 quanta or formation of a stable BH remnant (\verb+NP+=0). The results with \verb+GRAVITONLOSS+ $=1$ will be presented in a future report.

The CMS search for BH events looks at excess transverse energy w.r.t. SM background predictions \cite{exo-12-009}. The transverse energy $S_\mathrm{T}$ of an event is defined as the scalar sum of the transverse energies of all the final-state objects in excess of 50 GeV, i.e., jets, muons, electrons and photons satisfying the selection criteria discussed in Ref.\ \cite{exo-12-009}. The missing transverse energy is defined as the magnitude of the vector sum of the transverse momenta of all the final-state objects. If it is greater than 50 GeV, the missing transverse energy is added to $S_\mathrm{T}$. The event multiplicity $N$ is defined as the number of final-state objects which are used to calculate $S_\mathrm{T}$. BH events are expected to have high multiplicities.

\section{Results}

Lower bounds on $M_*$ and $M_\mathrm{min}$ are derived by evaluating the partial cross section $\sigma(S_\mathrm{T}>S_\mathrm{T}^\mathrm{min})$  for events whose $S_\mathrm{T}>S_\mathrm{T}^\mathrm{min}$ and whose multiplicities are greater than some chosen value, given by,
\begin{equation}
\sigma(S_\mathrm{T}>S_\mathrm{T}^\mathrm{min})=k\cdot\sigma_{pp\rightarrow\mathrm{BH}},
\end{equation}
where $S_T^\mathrm{min}$ is the minimal transverse energy chosen, and $k$ is 
\begin{equation}\label{factork}
 k=\frac{\mathrm{Num.\;of\; events\; with\; }S_\mathrm{T}>S_\mathrm{T}^\mathrm{min}}{\mathrm{Total\; Num.\; of\; events}}.
\end{equation}
The behavior of the total cross section $\sigma_{pp\rightarrow\mathrm{BH}}$ as a function of $M_*$ and $M_\mathrm{min}$ (i.e., \verb+XMIN+) follows from Eqs.(\ref{rs}, \ref{sigmatt}). There are 3 factors: 1) The BD cross section $\sigma_\mathrm{BD}$ is inversely proportional to a power of $M_*$; 2) Since \verb+XMIN+ appears as a lower limit of integration in the total cross section $\sigma_{pp\rightarrow \mathrm{BH}}$, the greater \verb+XMIN+, the smaller $\sigma_{pp\rightarrow \mathrm{BH}}$ is at a fixed $M_*$; 3) In addition, the PDFs fall off rapidly at high $Q$. Taking into account all these factors, $\sigma_{pp\rightarrow \mathrm{BH}}$ is expected to decrease as $M_*$ (\verb+XMIN+) increases  at fixed \verb+XMIN+ ($M_*$). The ratio $k$ in Eq.(\ref{factork}) can be estimated by integrating over the spectra of visible final state particles over the range $S_\mathrm{T}> S_\mathrm{T}^\mathrm{min}$ and then averaging over all final state particles. $k$ is an increasing function of the Hawking temperature, which, in turn, is a monotonically increasing function of $M_*$. Thus, as $M_*$ increases, the graph of $k$ vs.\ $S_\mathrm{T}^\mathrm{min}$ flattens. Following the CMS collaboration  \cite{glandsberg}, we choose the signal acceptance to be 100\%. In summary, $\sigma(S_\mathrm{T}>S_\mathrm{T}^\mathrm{min})\times A$ is expected to decrease as either $M_*$ or \verb+XMIN+ increases. 

\begin{figure}[h]
\center
\includegraphics[height=7.5cm]{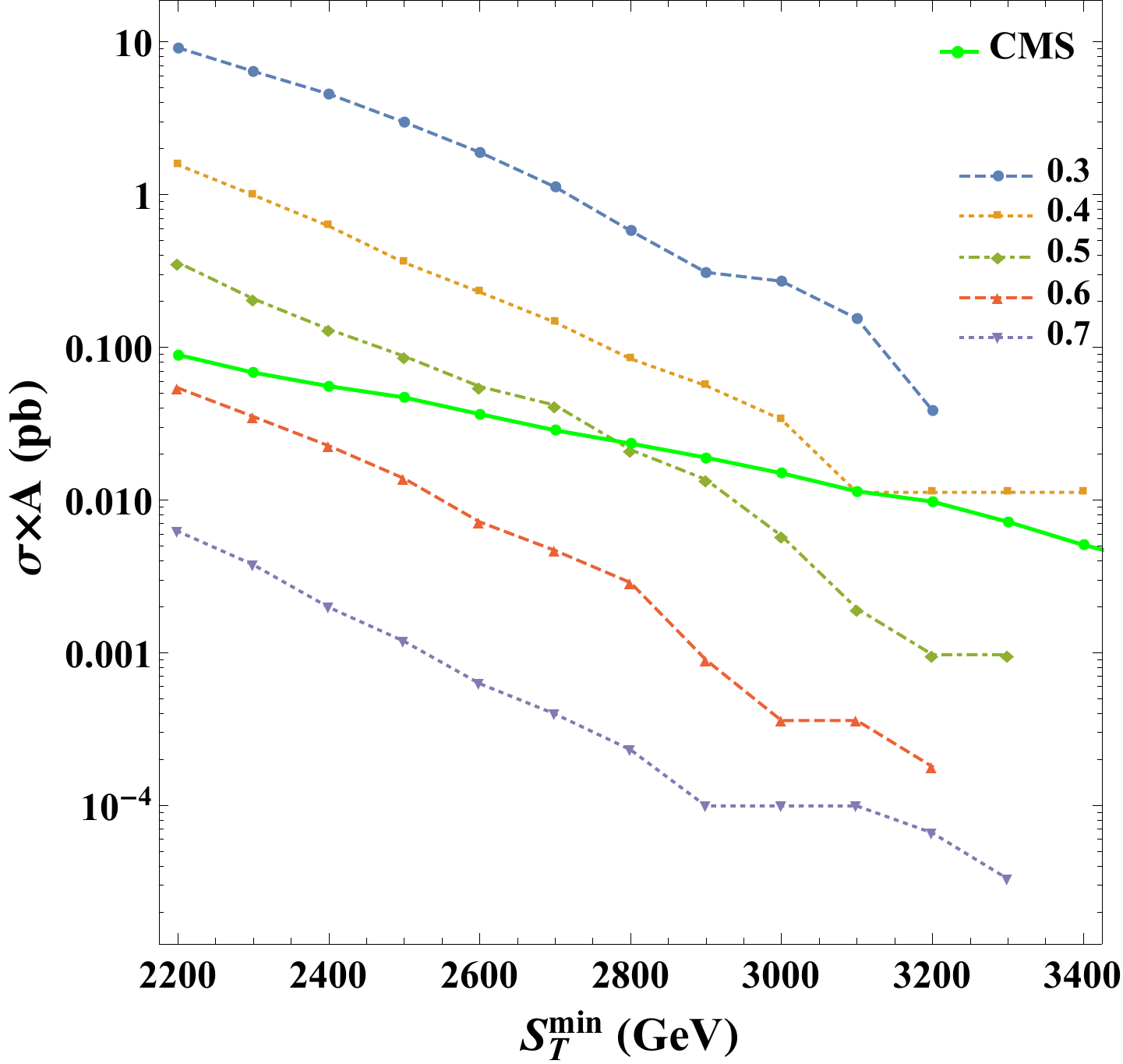}
\includegraphics[height=7.5cm]{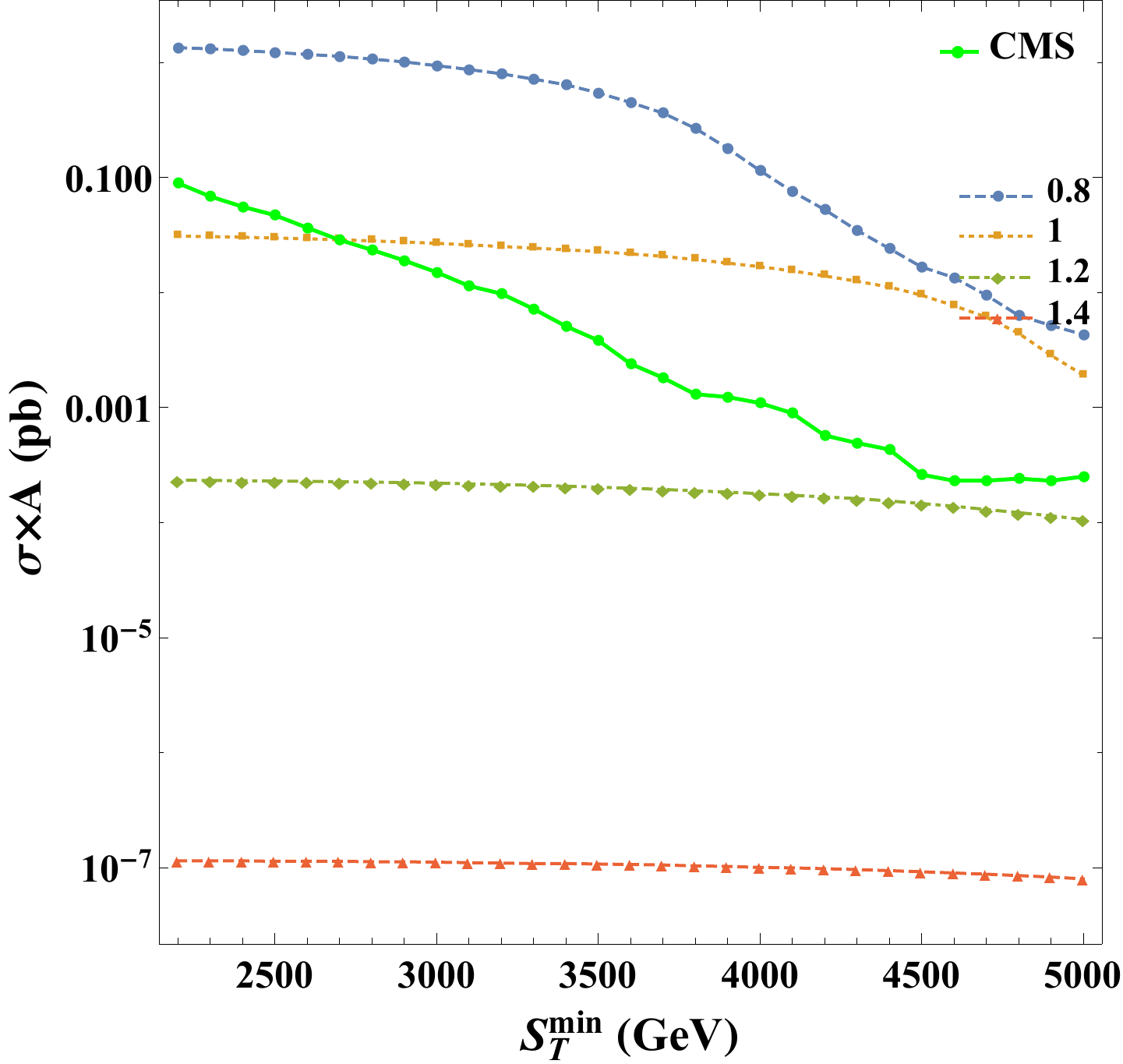}

\includegraphics[height=7.4cm]{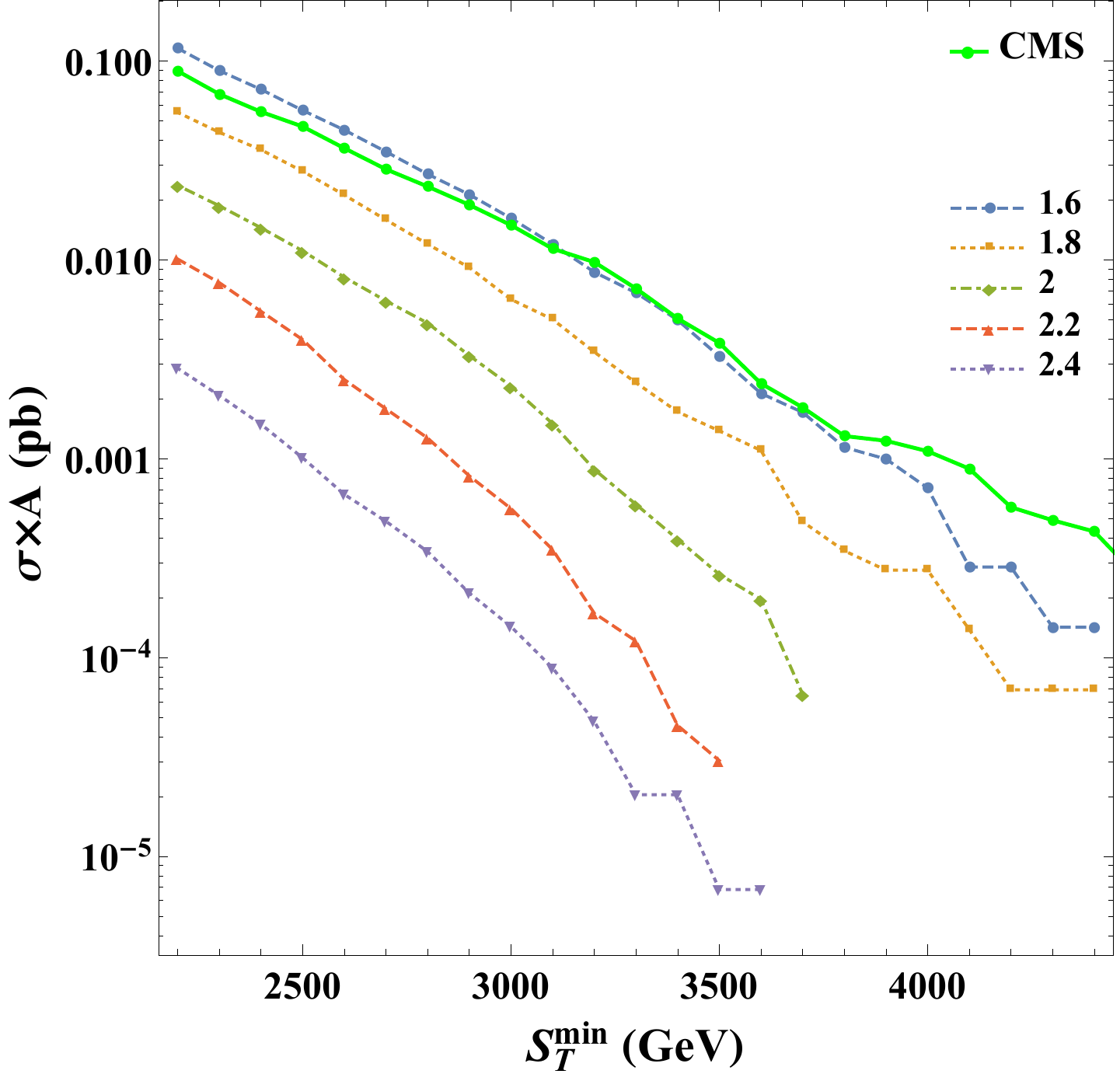}
\includegraphics[height=7.5cm]{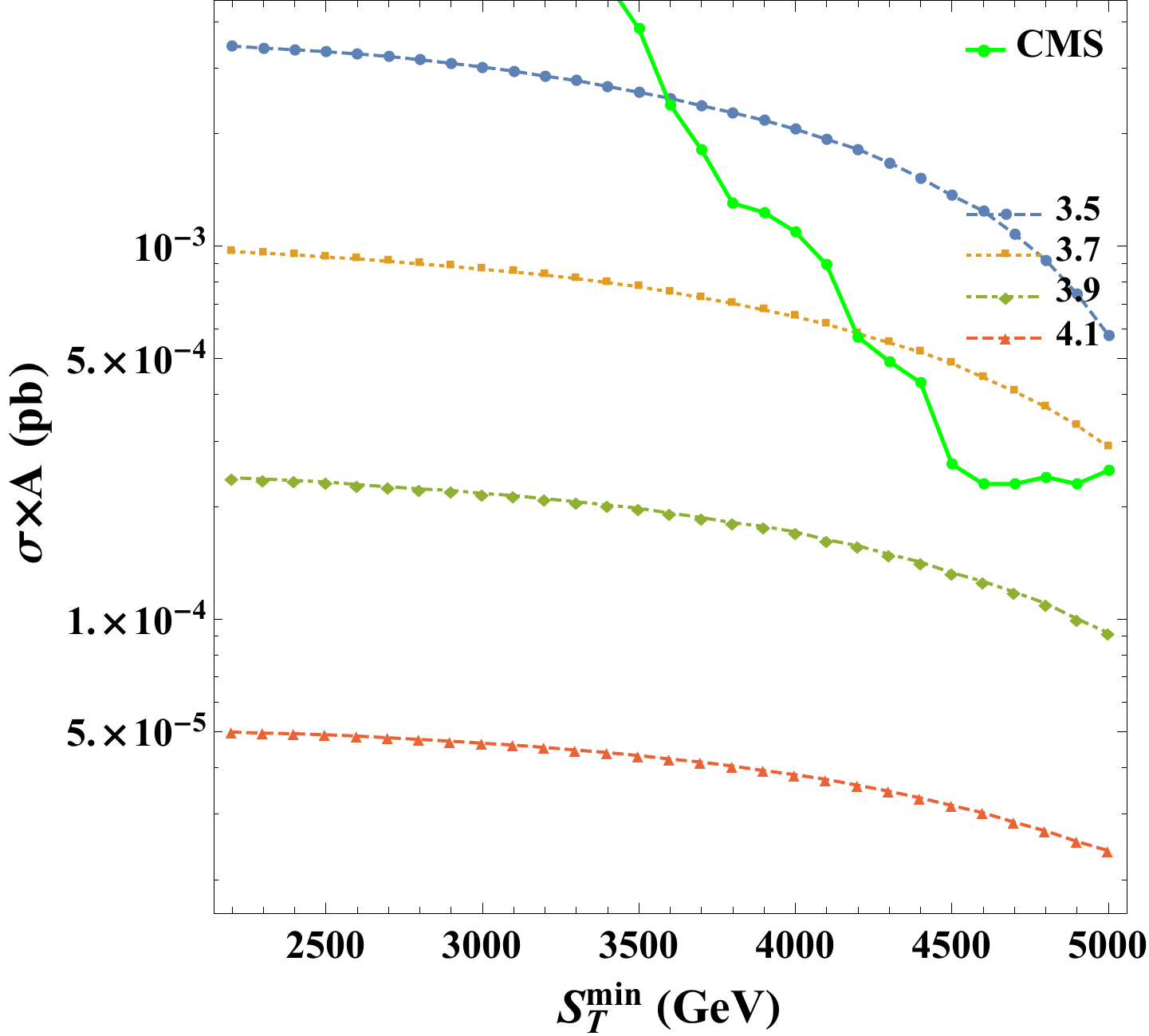}
\caption{\small $\sigma(S_\mathrm{T}>S_\mathrm{T}^\mathrm{min})\times A$ as a function of $M_*$ (Upper two graphs, numbers being $M_*$'s chosen in units of TeV) or XMIN (Lower two graphs, numbers being XMIN's) at NP = 0 (Right two graphs) and NP = 4 (Left two graphs). The model-independent 95\% CL experimental upper limits for counting experiments from CMS Collaboration are also shown. The multiplicity is $N\ge 3$.}\label{spectra}
\end{figure}

The partial cross section $\sigma(S_\mathrm{T}>S_\mathrm{T}^\mathrm{min})$ as a function of $M_*$ is obtained by running CATFISH with fixed \verb+NEXTRADIM+, \verb+GRAVITONLOSS+, \verb+XMIN+ = \verb+QMIN+, \verb+NP+.
The results are shown by the upper two graphs in Figure (\ref{spectra}) for \verb+NEXTRADIM+ = 3,  \verb+XMIN+ = \verb+QMIN+ = 5, \verb+NP+ = 0, 4. The lower two graphs in Figure (\ref{spectra}) display  $\sigma(S_\mathrm{T}>S_\mathrm{T}^\mathrm{min})$ as a function of \verb+XMIN+ for $M_*=1.5$ TeV, \verb+NEXTRADIM+ = 5, and \verb+NP+ = 0, 4.  As expected, Figure (\ref{spectra}) shows that the cross section decreases as $M_*$ and \verb+XMIN+ increase.

%
Figure (\ref{spectra}) can be used to determine bounds on $M_*$ and \verb+XMIN+ by comparing the simulated $\sigma(S_\mathrm{T}>S_\mathrm{T}^\mathrm{min})$ with the experimental limits (blue curves in the graphs). For example, the upper right plot shows that the lower limit on $M_*$ lies in the range 1.0 TeV$-$1.2 TeV, and the bottom right plot shows that the lower limit on \verb+XMIN+ lies in the range 3.7$-$3.9. Multiplying this range  by $M_*=1.5$ TeV, we obtain the lower limits on $M_\mathrm{min} = 5.55$ TeV$-$5.85 TeV.

We run CATFISH over a large range of parameter space as discussed in Appendix \ref{appa}. Figure (\ref{mslmts}) shows the exclusion region for $M_*$. As expected, the lower limit $M_{*, \, \mathrm{exp}}$ is a decreasing function of \verb+XMIN+. The value of $M_{*, \, \mathrm{exp}}$ does not strongly depend on \verb+NP+, as long as \verb+NP+ $ \ne 0$. If \verb+NP+ = 0, the bound on $M_{*, \, \mathrm{exp}}$ becomes much smaller. This is due to the high transverse momentum of the BH remnant, which contributes to the missing energy.  The lower limits on $M_*$  set upper bounds on $M_\mathrm{min}$ (\verb+XMIN+). As experimental data exclude values of $M_* \lesssim 1$ TeV\cite{catfish}, our results for \verb+NP+ $=$ 0 set an upper limit \verb+XMIN+ $\lesssim$ 2.5. \verb+NP+ $\ne 0$ results give the milder constraint, \verb+XMIN+ $\lesssim 6$. More experimental limits are shown in Table \ref{table1}, leading to more stringent constraints. For instance, CMS searches for events with an energetic jet and an imbalance in transverse momentum at $\sqrt{s}=8$ TeV \cite{exo-12-048} set the lower limit $M_*\sim 2.71$ TeV for $n=3$. The upper left panel in Figure (\ref{mslmts}) shows that the events with microscopic BHs decaying to remnants (\verb+NP+ = 0) are excluded, and the experimentally allowed range of \verb+XMIN+ is restricted to $1\sim 2$ for \verb+NP+ $\ne 0$. The lower limits of $M_*$ for $n=4,5,6$ do not exclude events with remnants as BH final products, but never the less, set strong constraints on the ranges where the semi-classical treatment is valid. 

\begin{figure}[h]

\center

\includegraphics[height=7.5cm]{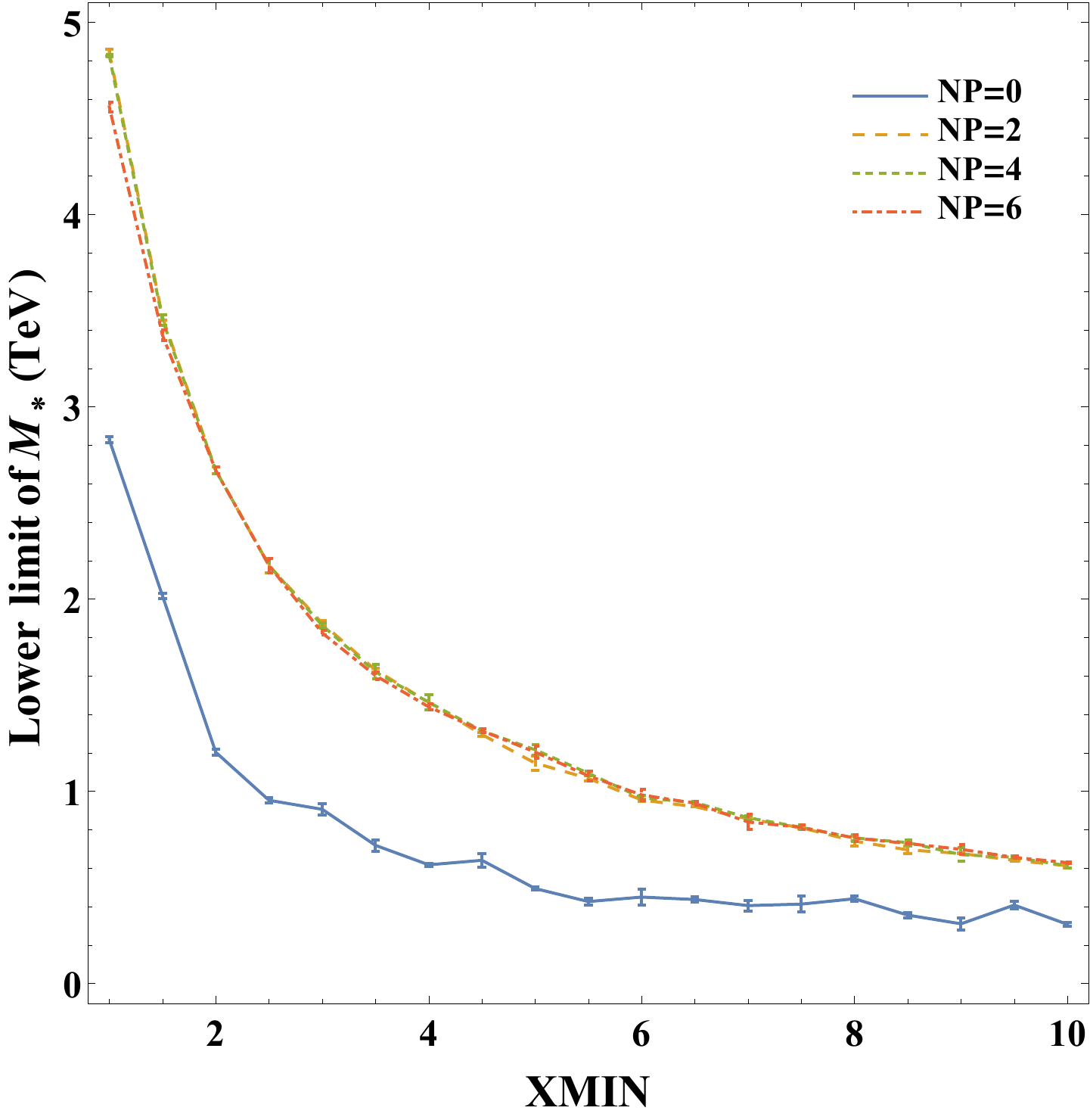}
\includegraphics[height=7.5cm]{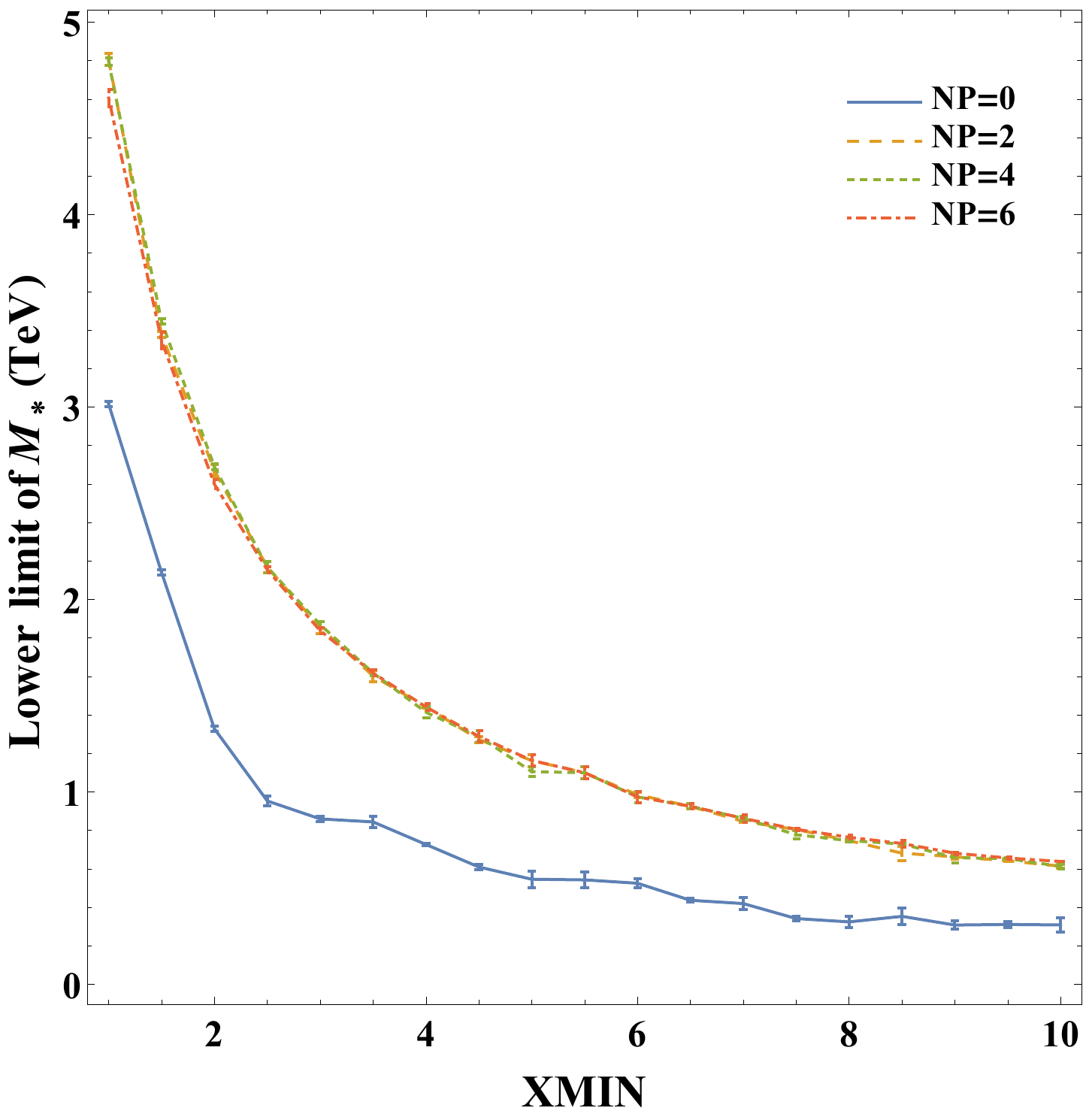}

\includegraphics[height=7.5cm]{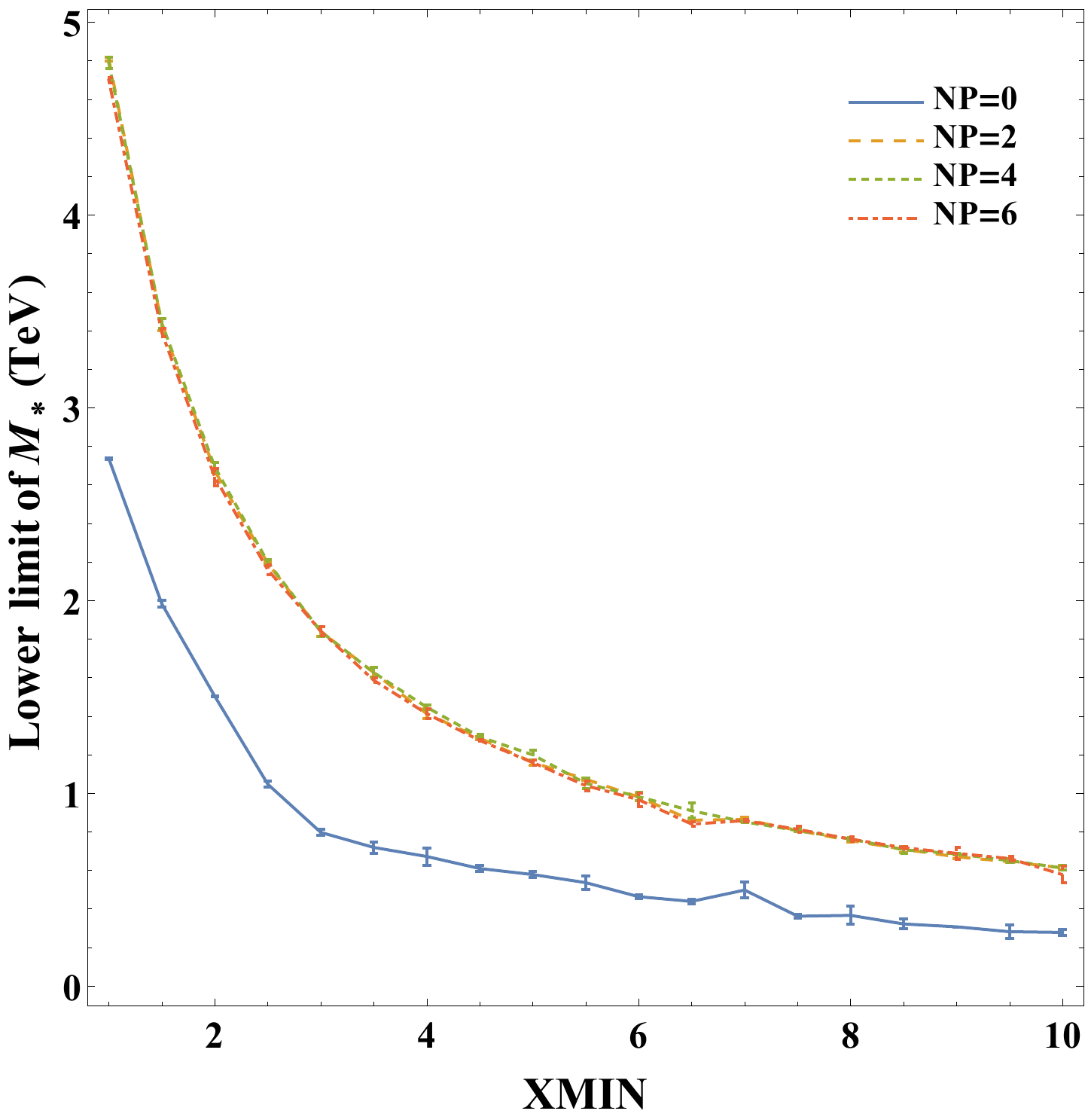}
\includegraphics[height=7.5cm]{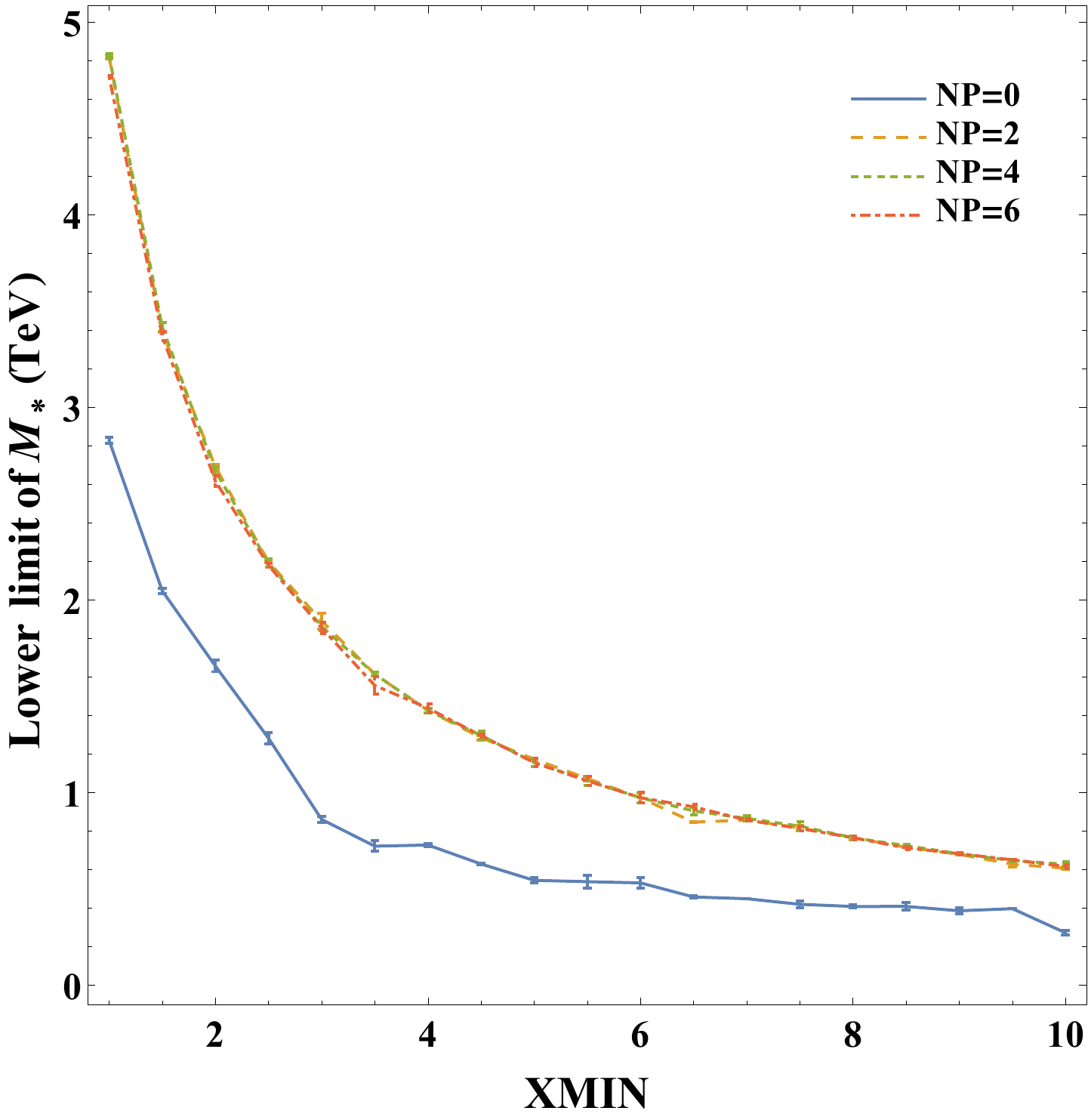}

\caption{\small Simulated lower limit on $M_*$ vs. XMIN as a function of NP and NEXTRADIM = 3 (top left), 4(top right), 5 (bottom left), and 6 (bottom right). }\label{mslmts}

\end{figure}

\begin{figure}[!h]

\center

\includegraphics[height=7.5cm]{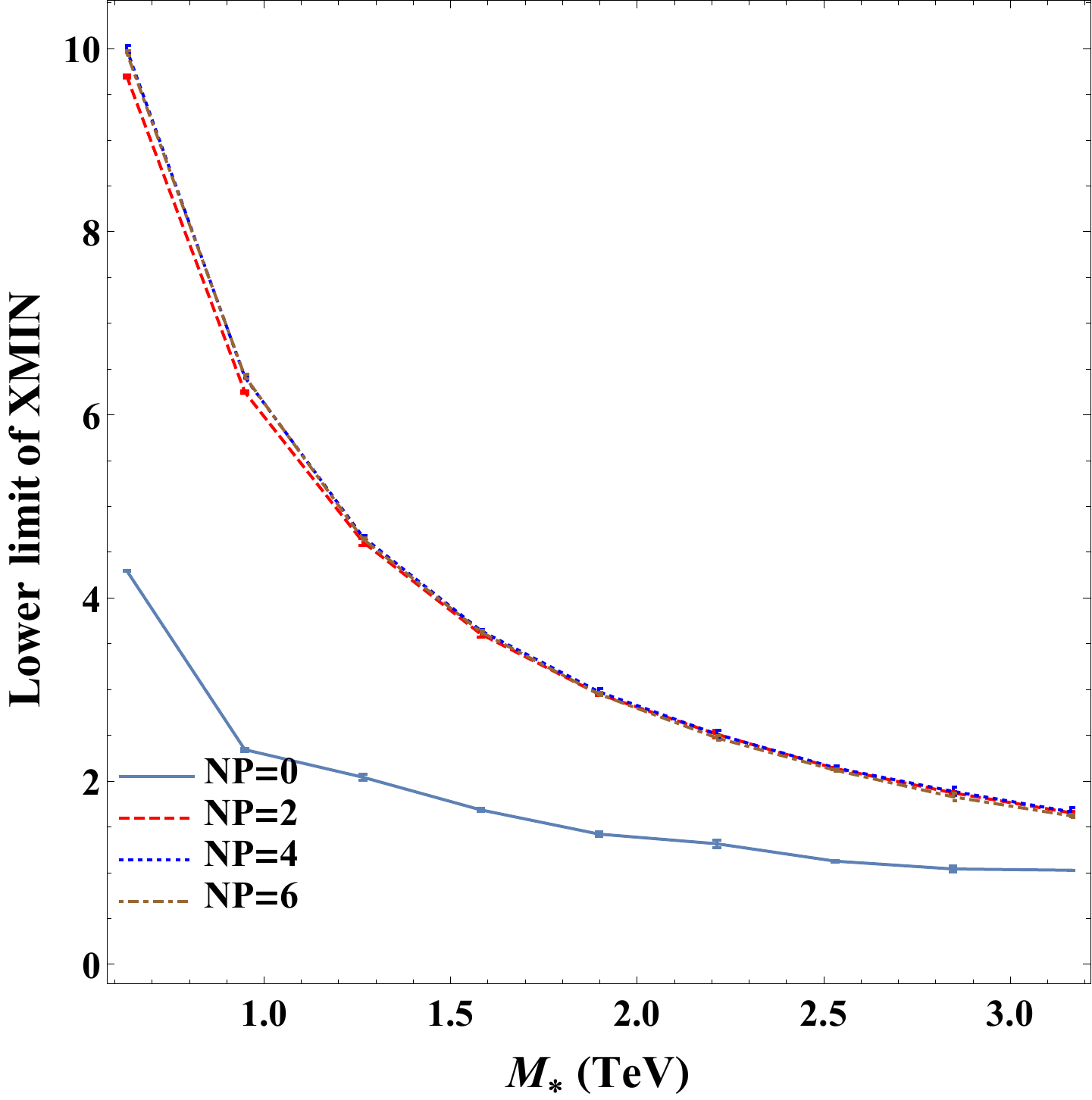}
\includegraphics[height=7.5cm]{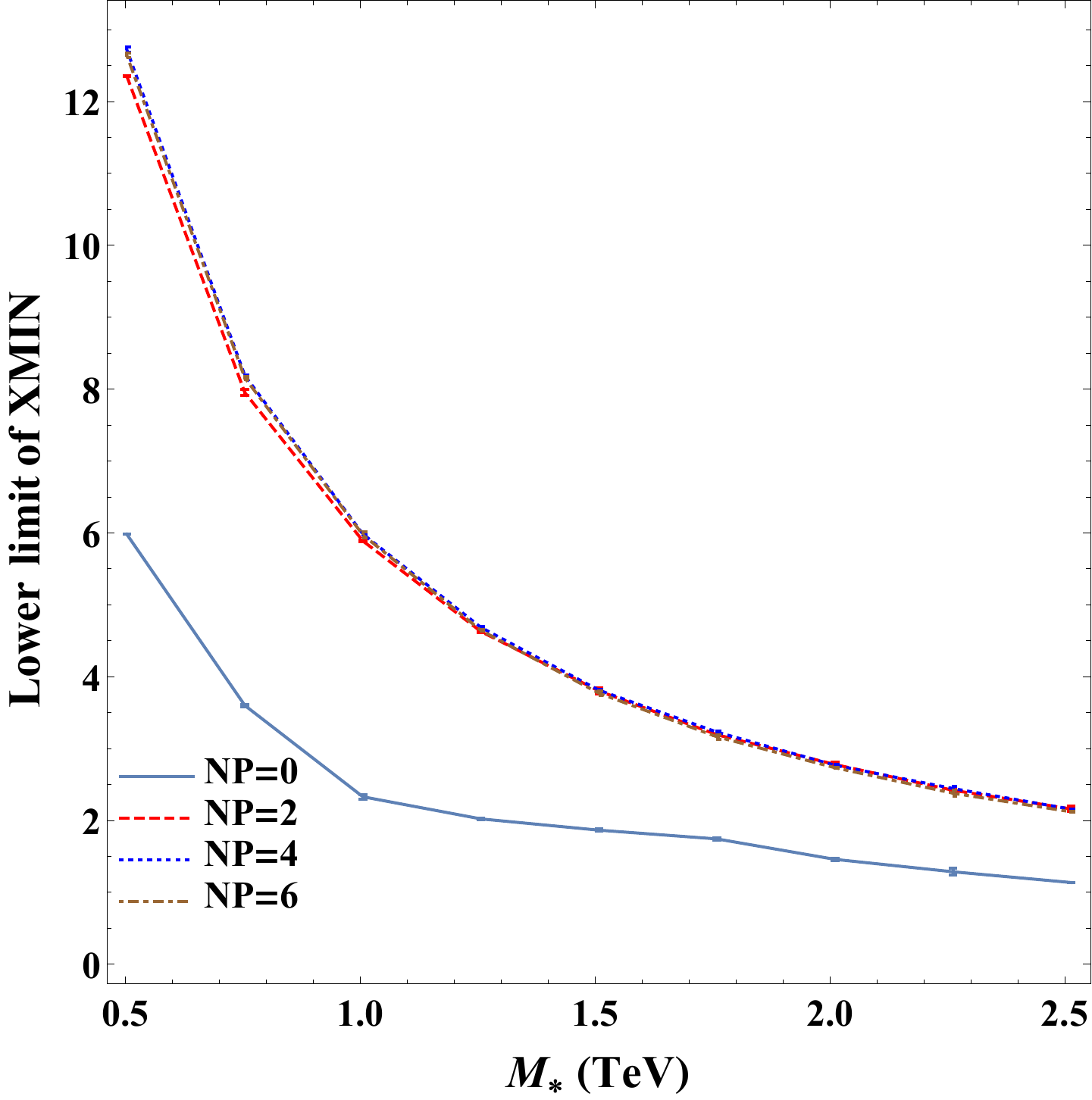}

\includegraphics[height=7.5cm]{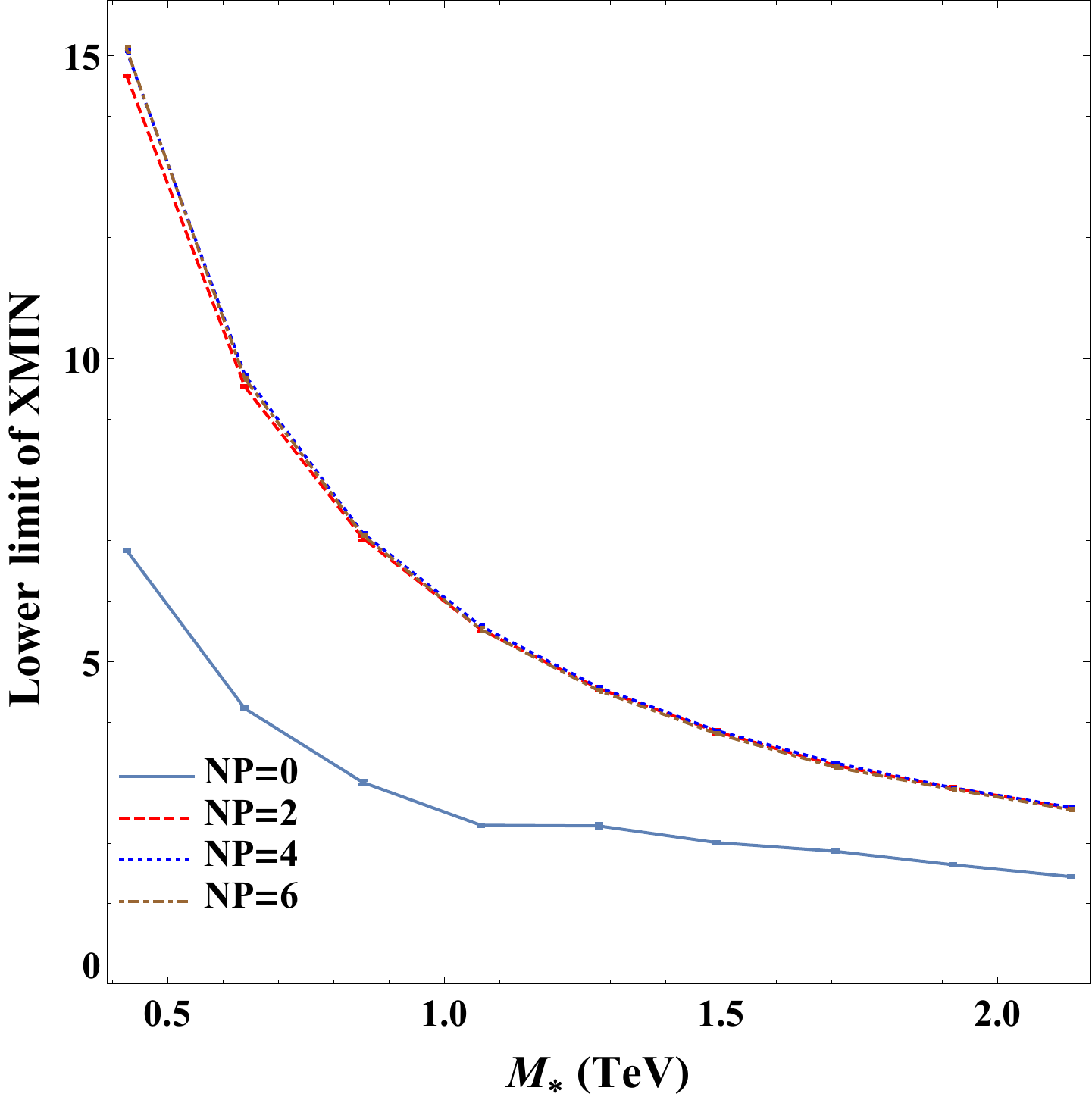}
\includegraphics[height=7.5cm]{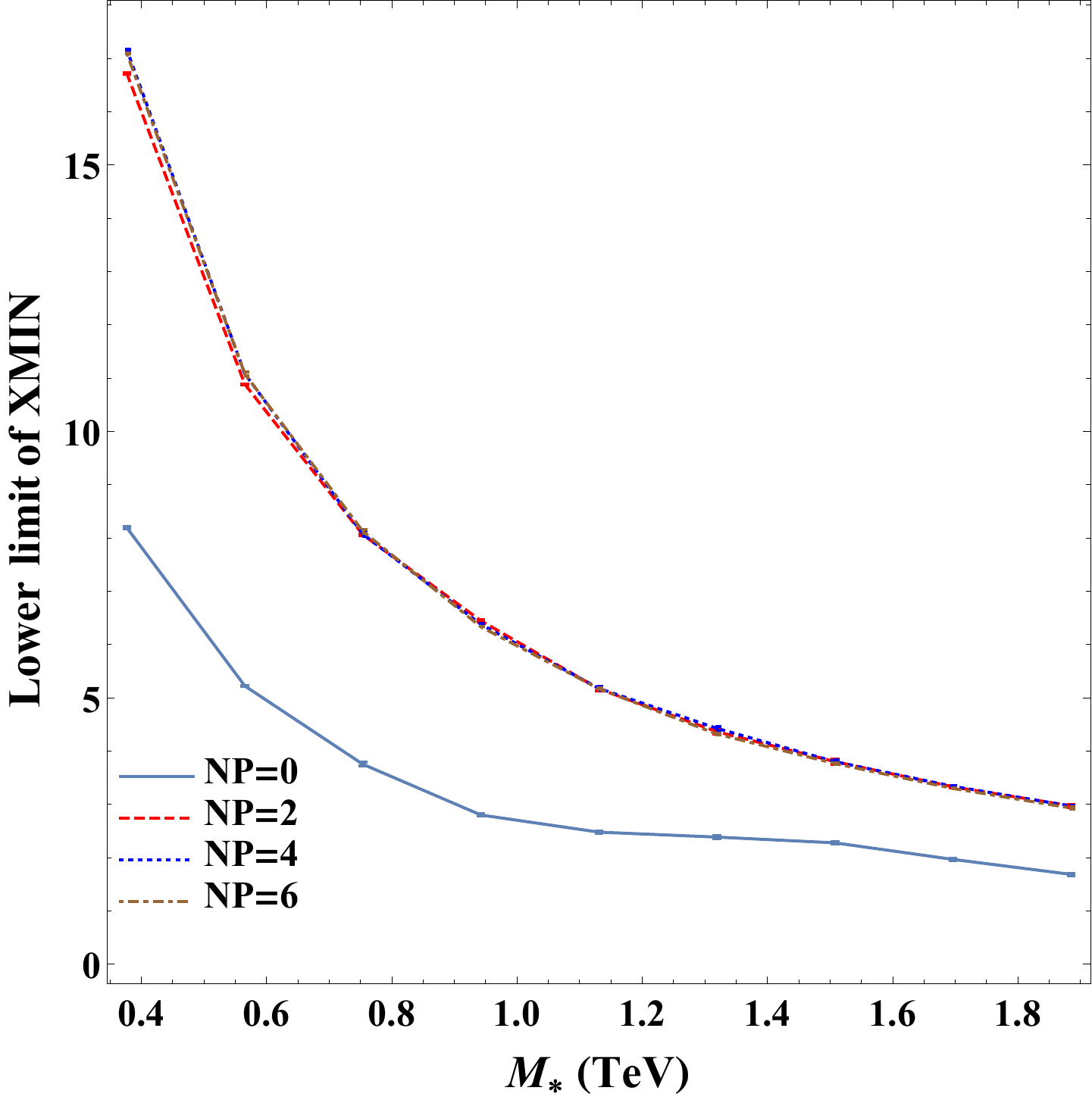}

\caption{\small Simulated lower limit on XMIN vs. $M_*$ as a function of NP and NEXTRADIM = 3 (top left), 4(top right), 5 (bottom left), and 6 (bottom right). }\label{xminlmts}

\end{figure}

Figure (\ref{xminlmts}) shows the exclusion region for \verb+XMIN+. As expected, the lower limit of \verb+XMIN+ is a decreasing function of $M_*$. Figure (\ref{xminlmts}) also shows that these limits do not depend on \verb+NP+ strongly when \verb+NP+ $\ne 0$, but become much smaller at \verb+NP+ = 0. This figure can be combined with Figure (\ref{mslmts}) to constrain $M_*$ further.  For example, if there are 3 extra dimensions, and a BH decays into 2 quanta in the quantum phase, the dashed curve (\verb+NP+ = 2, red onlne) in the upper left plot shows that \verb+XMIN+ $\gtrsim 4$ at $M_*\sim 1.5$ TeV. At the same time, the dashed curve  in the upper left plot ($n=3$) of Figure (\ref{mslmts}) indicates that if \verb+XMIN+ = 4, $M_* \gtrsim 1.5$ TeV. This shows that the lower limits of $M_*$ are consistent with those of \verb+XMIN+. In Figure (\ref{xminlmts}), the ranges of $M_*$ were chosen in order to compare the CATFISH results with those of BlackMax and CHARYBDIS2 from Ref.'s \cite{exo-12-009, atlasjhep}. 

\begin{figure}[!h]

\center

\includegraphics[height=7.5cm]{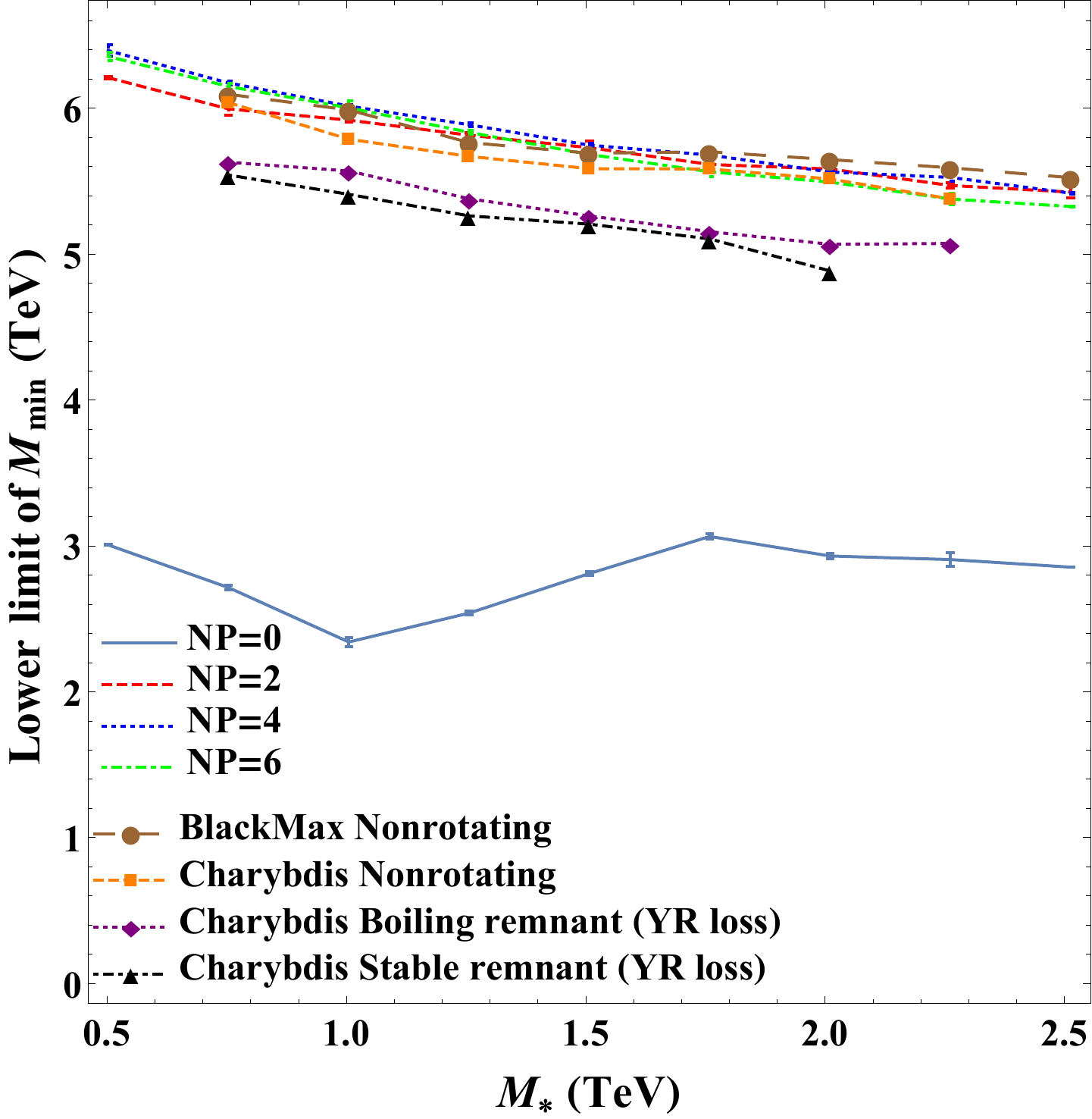}
\includegraphics[height=7.5cm]{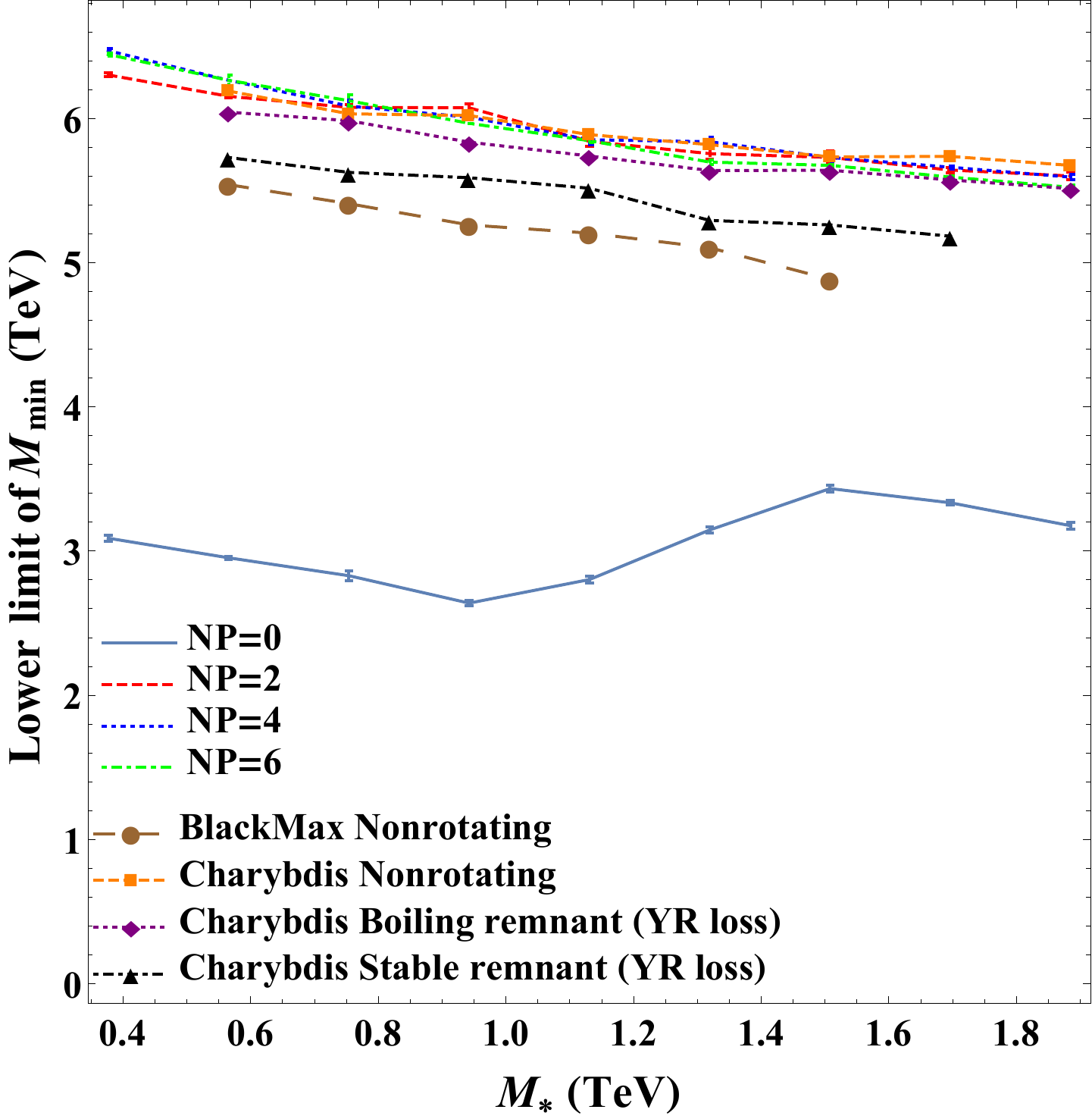}

\caption{\small Comparison of the predictions on lower limits of $M_\mathrm{BH}^\mathrm{min}$ from CATFISH with those from BlackMax and CHARYBDIS2 at NEXTRADIM = 4 (left) and 6 (right). The results of BlackMax and CHARBDIS are extracted from Fig. 4 in Ref.\cite{exo-12-009}. }\label{com}

\end{figure}
\begin{figure}[!h]

\center

\includegraphics[height=7.5cm]{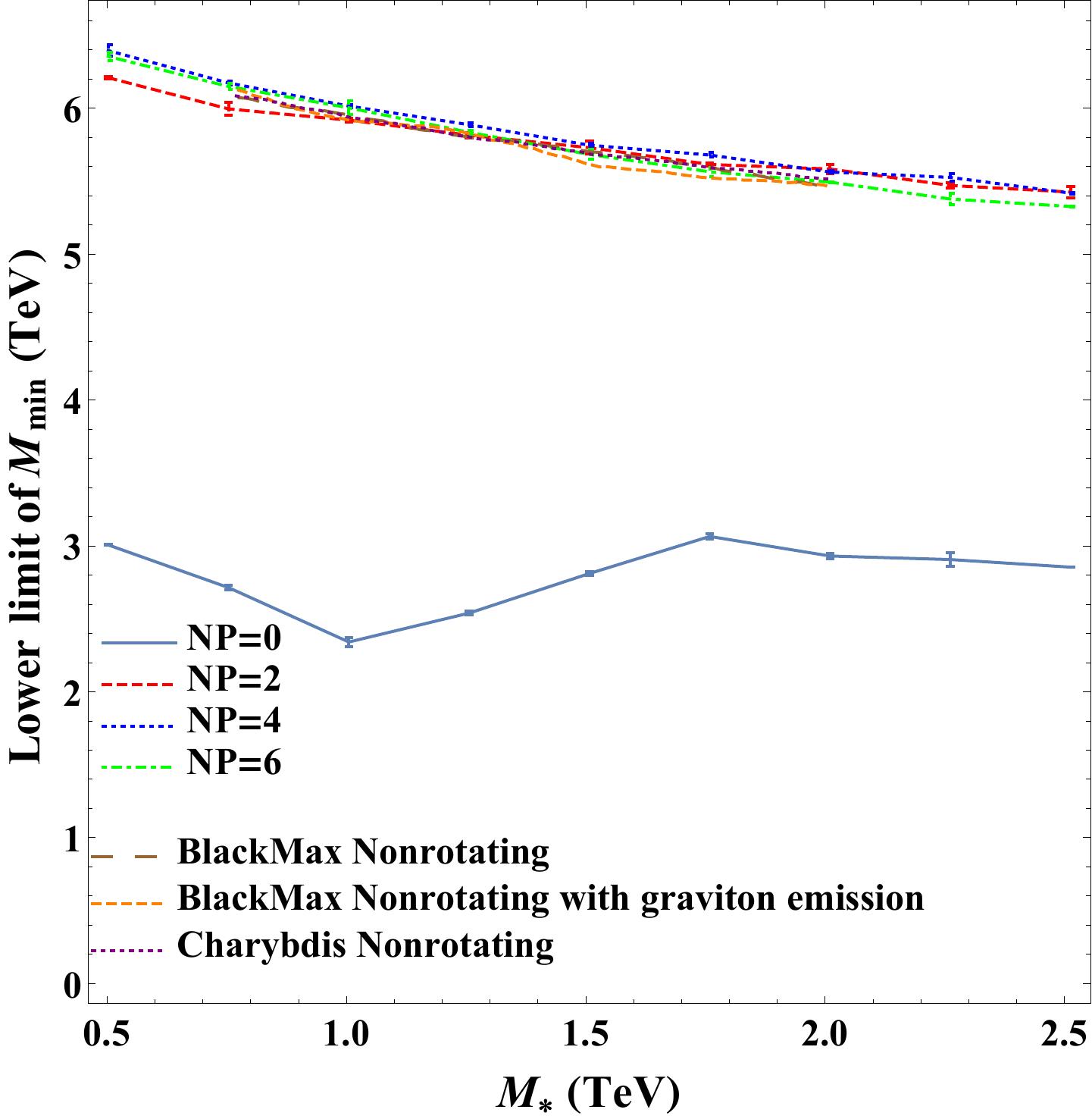}
\includegraphics[height=7.5cm]{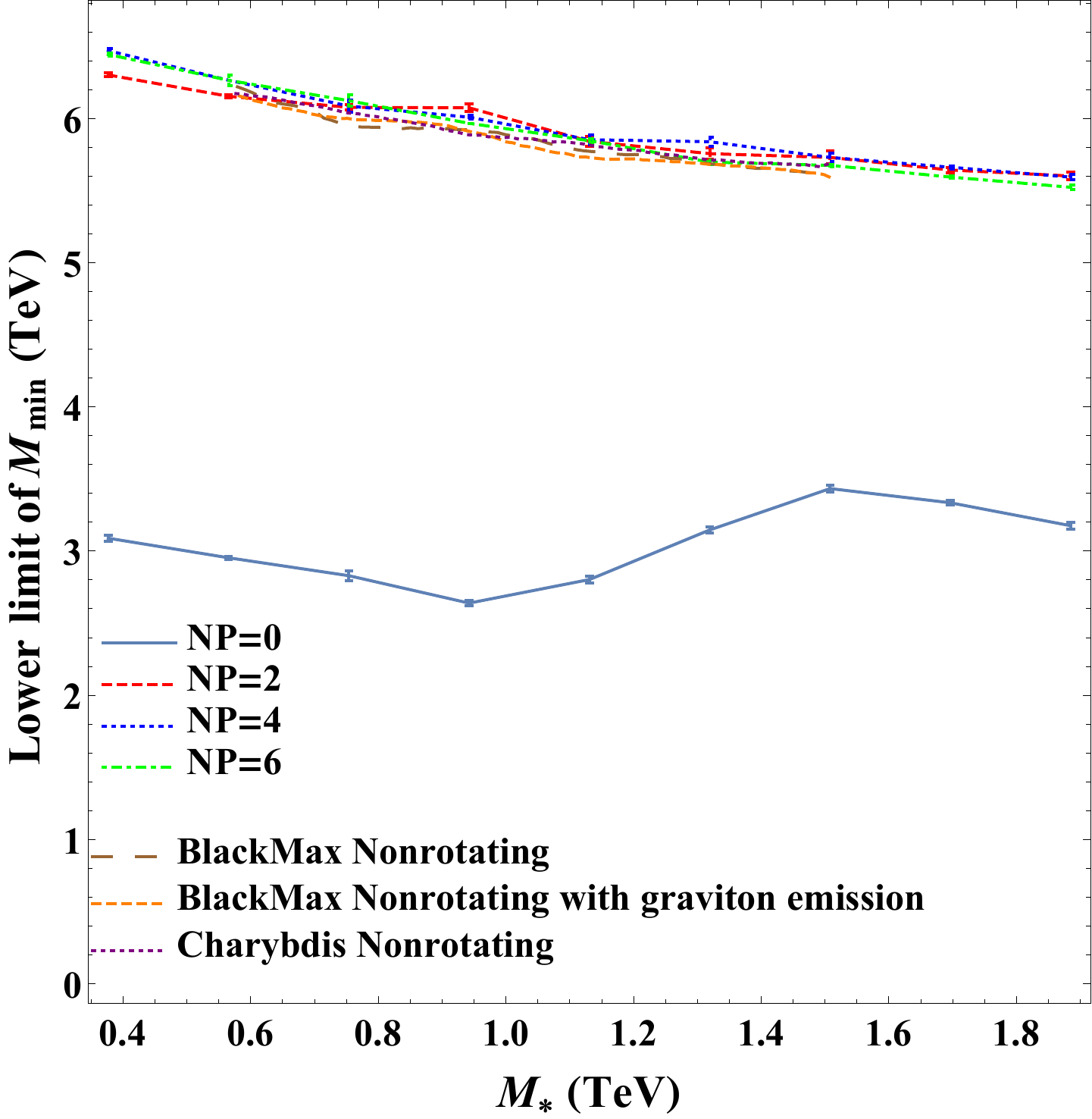}

\caption{\small Comparison of the predictions on lower limits of $M_\mathrm{BH}^\mathrm{min}$ from CATFISH with those from BlackMax and CHARYBDIS2 at NEXTRADIM = 4 (left) and 6 (right). The results of BlackMax and CHARBDIS are extracted from Fig.'s 8 and 10 in Ref.\cite{atlasjhep}. }\label{com2}

\end{figure}

Figure (\ref{com}) compares the lower limits of $M_\mathrm{min}$ predicted by CATFISH with those from BlackMax and CHARYBDIS2 done by CMS Collaboration \cite{exo-12-009}. It shows that as long as  \verb+NP+ = 0, CATFISH's limits are much smaller than those of BlackMax and CHARYBDIS2. When \verb+NP+ $\ne0$, the situation is more complicated. CATFISH agrees with BlackMax (nonrotating BH model) very well when $n=4$, but gives higher limits than BlackMax if $n=6$. CATFISH gives limits similiar to those of CHARYBDIS2 (nonrotating BH model) for $n=4,6$. While CHARYBDIS2's stable remnant model with YR loss predicts smaller limits than CATFISH at $n=4,6$, CHARYBDIS2's boiling remnant model with YR loss gives smaller limits at $n=4$, and similar limits to CATFISH at $n=6$.  Figure (\ref{com2}) shows the comparison between CATFISH predictions with those from ATLAS Collaboration \cite{atlasjhep}. CATFISH agrees with BlackMax and CHARYBDIS2 when \verb+NP+ $\ne0$, but predicts much smaller limits when \verb+NP+ = 0. The similarity displayed in the two figures is due to the fact that the 3 generators incorporate the same basic physics of microscopic BH formation and decay, but they differ from each other in the implementation of the quantum phase and/or  taking graviton energy loss at formation into account.

\section{Conclusions}

In this work, lower limits on the fundamental Planck scale $M_*$ and minimal BH mass $M_\mathrm{min}$ at formation have been obtained in a vast parameter space, using experimental upper limits on the partial production cross section of microscopic BHs \cite{exo-12-009}. Various models implemented in CATFISH have been explored and set different limits. In general, BH remnant models give milder constraints than non-remnant models. The predicted lower limits constrain the ADD model but do not exclude it. 
Future investigations will focus on performing a similar analysis to that carried out in the present paper with the additional feature of graviton energy loss during BH formation (\verb+GRAVITONLOSS+ = 1).  Another refinement of the models is to include the effects of the generalized uncertainty principle \cite{gup}. The steps outlined above for the simulation of microscopic BH events can also be carried out for string balls, string resonances and other exotic particles. 

\section{Acknowledgements}

We wish to thank C. Henderson and P. Rumerio for constructive discussions, Greg Landsberg for helpful comments on the CMS data, and the CMS Exotica conveners for providing numerical cross section limits. We also gratefully acknowledge RC2 \cite{rc2} at the University of Alabama and Alabama Supercomputer Authority \cite{asa} for providing the computing infrastructure essential to our analyses.
This research was supported in part by the DOE under grant DE-FG02-10ER41714.

\newpage

\appendix

\section{Appendix A: Procedure for determining $M_*$ and $M_\mathrm{min}$ limits}\label{appa}

The primary goal of the work is to determine the limits on the $D$-dimensional fundamental Planck scale $M_*$, or equivalently, the reduced Planck scale $M_D$ used in Ref. \cite{exo-12-009}, for different  \verb+NEXTRADIM+ (=3, 4, 5, 6), \verb+GRAVITONLOSS+ (=0), \verb+XMIN+ (= \verb+QMIN+) and \verb+NP+ (=0, 2, 4, 6). The model-independent 95\% C.L. upper cross section limits for counting experiments (Figures 6 and 7 in Ref. \cite{exo-12-009}) were used. 
CATFISH accepts only \verb+XMIN+ $\ge$ 1, so the lower limit of \verb+XMIN+ is 1 and its upper limit is determined by noticing that the mass of the BH should be less than the center of mass energy of the LHC,
\begin{equation}\label{condition1}
\verb+XMIN+\times \verb+MSTAR+\le 8 \;\textrm{TeV}\,. 
\end{equation}
The stepsize of \verb+XMIN+ is set to 0.5. We use a bisection method to find the limits on the $D$-dimensional fundamental Planck scale:

Step 1. The possible range of $M_*$ is determined by fixing all other parameters. The lower limit is $m_0=1$ TeV and the upper limit, $M_0$, satisfies Eq.\ (\ref{condition1}) and the condition 
\begin{equation}
\sigma\cdot L \ge \frac{1}{N_\textrm{run}},\label{cond}
\end{equation}
where $\sigma$ is the cross section of the production, $L=12 \;\textrm{fb}^{-1}$ is the LHC integrated luminosity, and $N_\textrm{run}$ is the number of events of each run. In our simulations we choose $N_\textrm{run}=10^4$.

Step 2. CATFISH is run with \verb+MSTAR+=$M_1=(m_0+M_0)/2$ and the simulation output is compared with CMS results \cite{exo-12-009}.

Step 3. If $M_1$ is allowed by experimental data, i.e., the simulated cross section (times the detector acceptance $A$) for BH production is too small, $M_*$ must be smaller than $M_1$. CATFISH 
is run with \verb+MSTAR+=$M_2=(m_0+M_1)/2$. Otherwise, CATFISH is run with \verb+MSTAR+=$M_2=(M_1+M_0)/2$.

Step 4. Step 3 is repeated $i$ times until $|M_i-M_{i-1}| < \Delta M$, where $M_i$ is the result of the $i$-th simulation and $\Delta M$ is the required precision, $\Delta M=0.1$ TeV.

Step 5. The more likely value of $M_*$ is then determined as 
\[
M_{*,\,\mathrm{exp}}=\frac{M_i+M_{i-1}}{2}\,. 
\]
The error on $M_{*,\,\mathrm{exp}}$ is $\Delta M_{*,\,\mathrm{exp}}=p|M_i-M_{i-1}|$, where $p$ is determined at a given confidential level (C.L.) for the interval $[\mathrm{Min}(M_i,\, M_{i-1}), \mathrm{Max}(M_i, \, M_{i-1})]$,
\[
p=\frac{1-\mathrm{C.L.}/100}{2}. 
\]
The lower limit on the Planck mass, $M_{*, \, \mathrm{exp}}$, is determined for a given choice of \verb+NEXTRADIM+, \verb+GRAVITONLOSS, XMIN+ (= \verb+QMIN+) and \verb+NP+. Different sets of these parameters are chosen and Steps 1-5 are repeated to determine $M_{*, \, \mathrm{exp}}$ as function of the parameters. Similarily, the lower limit of $M_\mathrm{min}$ or \verb+XMIN+ can be obtained for different choices of \verb+MSTAR+, \verb+NEXTRADIM+, \verb+GRAVITONLOSS+ (=0) and \verb+NP+.

\end{document}